\documentclass[preprint,showpacs,showkeys,preprintnumbers,amsmath,amssymb]{revtex4}
 \usepackage{dcolumn}
 \usepackage{bm}

\begin{document}

 \title{ Hawking Radiation of Fermionic Field and
  Anomaly in 2+1 Dimensional Black Holes}

 \author{Ran Li}
 \thanks{Electronic mail: liran05@lzu.cn}
 \author{Shi Li}
 \thanks{Electronic mail: lsh08@lzu.cn}
 \author{Ji-Rong Ren}
 \thanks{Corresponding author. Electronic mail: renjr@lzu.edu.cn}

 \affiliation{Institute of Theoretical Physics, Lanzhou University, Lanzhou, 730000, Gansu, China}

 \begin{abstract}

 The method of anomaly cancellation
 to derive Hawking radiation
 initiated by Robinson and Wilczek is applied to
 2+1 dimensional stationary black holes.
 Using the dimensional reduction technique,
 we find that the near-horizon physics for
 the fermionic field in the background
 of the general 2+1 dimensional stationary black hole
 can be approximated by an infinite collection of
 two component
 fermionic fields in 1+1 dimensional spacetime background
 coupled with dilaton field and $U(1)$
 gauge field. By restoring the gauge invariance
 and the general coordinate covariance for the reduced two
 dimensional theory,
 Hawking flux and temperature of black hole are obtained.
 We apply this method to two types of black holes in three dimensional
 spacetime, which are BTZ black hole in Einstein gravity
 and a rotating black hole
 in Bergshoeff-Hohm-Townsend (BHT) massive gravity.

 \end{abstract}



 \maketitle

 \section{Introduction}

 Using the techniques of quantum field theory in curved
 spacetime,
 Hawking made the discovery\cite{hawk,hawking} that black hole
 can radiate from its event
 horizon with a purely thermal spectrum at the
 temperature $T=\frac{\kappa}{2\pi}$($\kappa$ is
 the surface gravity of the black hole),
 which leads to a remarkable connection
 between thermodynamics, quantum mechanics and gravity
 \cite{ben}.
 Due to the significance of Hawking radiation and
 the difficulties existing in applying the Hawking's
 original method to more complicated spacetime backgrounds,
 several derivations of Hawking radiation have been
 proposed in literatures, including the so-called
 Damour-Ruffini method\cite{damour}, trace anomaly method
 \cite{trace}, quantum tunneling method\cite{parikh} and
 gravitational anomaly method\cite{graanomaly1,graanomaly}.

 The recent proposal of deriving Hawking radiation
 via gravitational and gauge anomalies
 proposed by Wilczek and his collaborators
 \cite{graanomaly1,graanomaly}
 has attracted a lot of interests.
 This rejuvenates the interest of investigation
 of Hawking radiation. For various types of
 black holes, these investigations have been
 carried out \cite{investigations}.
 In fact, the anomaly analysis can
 be traced back to Christensen and
 Fulling's early work \cite{trace}, in
 which they suggested that there
 exists a relation between the
 Hawking radiation and anomalous
 trace of the field under the
 condition that the covariant
 conservation law is valid.
 Imposing boundary condition near the horizon,
 Wilczek et al showed that the
 Hawking radiation is just the
 cancel term of the gravitational
 anomaly of the covariant conservation
 law and gauge invariance.
 Their basic idea is that, near the horizon,
 a quantum field in a
 black hole background can be effectively
 described by an infinite
 collection of (1+1)-dimensional fields
 on $(t, r)$ space, where $r$ is
 the radial direction. One could
 treat the original higher
 dimensional field as a collection
 of two-dimensional quantum fields
 using the dimensional reduction.
 In this two-dimensional reduction,
 because all the ingoing modes can not
 classically affect physics outside the horizon,
 the two-dimensional effective action
 in the exterior region becomes anomalous with
 respect to gauge or general coordinate symmetries.
 To cancel the anomaly, they found the Hawking flux
 is universally determined only
 by the value of anomalies at the horizon.
 It is also interesting to generalize the
 above idea to derive the high-spin flux from a black hole
 through the conformal symmetry in reduced $(1+1)$-dimensional
 background\cite{high}. There are also some interests to
 investigate the relationship between
 quantum anomaly and tunneling method\cite{tun}.

 Originally, in order to apply the anomaly
 cancellation method, only scalar field
 is considered in the dimensional reduction
 process. However, the anomaly formalism
 can also be used to investigate
 the Hawking radiation of vector field\cite{vector} and spinor
 field\cite{spin}. In \cite{vector}, it is explicitly shown
 that the theory of an electromagnetic field on $d$-dimensional
 spherical black holes can be reduced to one of an infinite number
 of massive complex scalar fields on
 2-dimensional spacetime,
 for which the usual anomaly cancellation
 method is available.
 For the Hawking radiation of spinor field\cite{spin},
 it turns out that the near-horizon physics for
 fermionic field in Kerr black hole can be
 approximated by an effective two dimensional field theory.
 Therefore, in order to verify the universality
 of the anomaly cancellation method
 in deriving Hawking radiation,
 it is interesting to discuss the
 Hawking radiation for other fields
 with nonzero spins.

 In the present work,
 we extend the gravitational anomaly
 method to investigate
 the Hawking radiation of fermionic field in
 the 2+1 dimensional stationary black hole.
 Using the dimensional reduction technique,
 we find that the near-horizon physics for
 the fermionic field in the background
 of the general 2+1 dimensional stationary black hole
 can be approximated by an infinite collection of
 two component
 fermionic fields in 1+1 dimensional spacetime background
 coupled with the dilaton field and the $U(1)$
 gauge field. By restoring the gauge invariance
 and the general coordinate covariance for the reduced two
 dimensional theory,
 the Hawking temperature and flux of black hole are obtained.
 We apply this method to two types of black holes in three dimensional
 spacetime, which are BTZ black hole\cite{BTZ} in Einstein gravity
 and a rotating black hole\cite{BHTso}
 in Bergshoeff-Hohm-Townsend (BHT) massive gravity\cite{BHT}.
 It turns out that the Hawking temperatures are recovered and they are
 consistent with the results obtained previously
 in literatures.

 This paper is organized as follows:
 In section 3, by performing the dimensional reduction
 for the fermionic field action in the general 2+1 dimensional
 stationary black hole background,
 we obtain the reduced $1+1$ dimensional effective theory and
 calculate the Hawking flux
 via anomaly equations near the horizon.
 In section 3 and 4, the method is applied
 to BTZ black hole in Einstein gravity
 and a rotating black hole
 in BHT massive gravity.
 The last section is devoted to conclusions and discussions.

 \section{Hawking radiation of fermionic field and anomalies}

 In this section, let us consider the general 2+1
 dimensional stationary black hole spacetime,
 of which the line element is given by\cite{cqg2+1}
 \begin{eqnarray}
 ds^2=-h(r)dt^2+\frac{1}{f(r)}dr^2+r^2(d\phi+N^\phi(r)dt)^2\;.
 \end{eqnarray}
 We assume that the event horizon is
 located at $r=r_H$. Meanwhile the metric functions
 satisfy the equations $h(r_H)=0$ and $f(r_H)=0$.
 The angular velocity of the black hole
 is given by $\Omega_H=-N^\phi(r_H)$. The surface gravity
 can be calculated from the standard formula
 $\kappa^2=-\frac{1}{2}k_{\mu;\nu}k^{\mu;\nu}|_{r_H}$,
 where the killing vector $k$ is given by
 \begin{eqnarray}
 k=\partial_t+\Omega_H\partial_\phi\;.
 \end{eqnarray}
 After some algebra, one can obtain
 \begin{eqnarray}
 \kappa=\frac{1}{2}\sqrt{\frac{f}{h}}h'\Big|_{r=r_H}\;,
 \end{eqnarray}
 where the prime $'$ denotes the derivative with respect
 to the coordinate $r$.
 For simplicity, we will consider the
 special case when
 $h(r) = f(r)g(r)$ and $g(r_H)\neq 0$.
 In this case, the surface gravity can be
 rewritten as $\kappa=\frac{1}{2}\sqrt{f'(r_H)h'(r_H)}$.
 The Hawking temperature
 given by the surface gravity is
 $T_H=\frac{\kappa}{2\pi}=\frac{1}{4\pi}\sqrt{f'(r_H)h'(r_H)}$.

 We consider the action for the fermionic field
 in this background
 \begin{eqnarray}
 S&=&\int d^3x \sqrt{-g} \bar{\psi}\gamma^a
 e_a^\mu D_\mu \psi\nonumber\\
 &=&\int d^3x \sqrt{-g} \psi^\dag \gamma^0\gamma^a
 e_a^\mu\left(\partial_\mu-\frac{1}{8}
 \omega_{bc\mu}[\gamma^b,\gamma^c]\right)\psi\;,
 \end{eqnarray}
 where the spin connection is given by
 $\omega_{ab\mu}=e_{a\nu}\nabla_\mu e_b^\nu$. It should be noted
 that the indices $a,b,c=0,1,2$ are flat while
 the indices $\mu,\nu=t,r,\phi$ are curved.
 $\gamma^a$ is the gamma matric in three dimensional
 flat spacetime.
 According to the line
 element (1), the tetrad field $e_a^\mu$ can be selected as
 \begin{eqnarray}
 e_0^\mu&=&(\frac{1}{\sqrt{h}},0,-\frac{N^\phi}{\sqrt{h}})\;,\nonumber\\
 e_1^\mu&=&(0,\sqrt{f},0)\;,\nonumber\\
 e_2^\mu&=&(0,0,\frac{1}{r})\;.
 \end{eqnarray}
 One can also define the spin connection
 with three flat indices as
 $\omega^a_{\;\;bc}=e_c^\mu\omega^a_{\;\;b\mu}
 =e_c^\mu e^a_{\nu}\nabla_\mu e_b^\nu$.
 After some algebra, one can obtain the non-vanishing components of
 spin connection as listed in the following
 \begin{eqnarray}
 \omega^1_{\;\;00}&=&\frac{\sqrt{f}h'}{2h}\;,\nonumber\\
 \omega^1_{\;\;02}&=&\omega^2_{\;\;01}=
 \omega^2_{\;\;10}=-\frac{r\sqrt{f}}{2\sqrt{h}}
 \frac{dN^{\phi}}{dr}\;,\nonumber\\
 \omega^2_{\;\;12}&=&\frac{\sqrt{f}}{r}\;.
 \end{eqnarray}
 When substituting the tetrad field and spin connection
 into the fermionic field action, one can obtain
 \begin{eqnarray}
 S=\int dtdrd\phi r\sqrt{\frac{h}{f}}
 \psi^\dag \gamma^0 &\Big\{&\frac{\gamma^0}{\sqrt{h}}
 (\partial_t-N^\phi\partial_\phi)+\gamma^1\sqrt{f}\partial_r
 +\frac{\gamma^2}{r}\partial_\phi\nonumber\\
 &&-\frac{1}{4}\left(\frac{\sqrt{f}h'}{h}\gamma^1
 +\frac{2\sqrt{f}}{r}\gamma^1-\frac{r\sqrt{f}}{\sqrt{h}}
 \frac{dN^{\phi}}{dr}\right)\Big\}\psi\;.
 \end{eqnarray}
 To proceed, one can define the tortoise coordinate as
 \begin{eqnarray}
 \frac{dr*}{dr}=\frac{1}{f(r)}\;.
 \end{eqnarray}
 When taking the near horizon limit
 $r\rightarrow r_H$,
 the metric function $f(r)\rightarrow 0$
 and $h(r)\rightarrow 0$.
 After ignoring the sub-leading contribution
 of the terms in the action, the action can be
 effectively simplified as
 \begin{eqnarray}
 S=\int dtdrd\phi r \sqrt{\frac{h}{f}}\psi^\dag
 \gamma^0\left[\frac{\gamma^0}{\sqrt{h}}
 (\partial_t-N^\phi\partial_\phi)+
 \frac{\gamma^1}{\sqrt{f}}(\partial_{r*}-\frac{fh'}{4h})\right]\psi\;.
 \end{eqnarray}
 In order to perform the integral over the angular $\phi$
 in the action, one can expand the two component spinor
 field $\psi$ in the following way
 $\psi(t,r,\phi)=\sum_m\psi_m(t,r)e^{im\phi}$.
 Note that $\psi_m(t,r)$ is also a two
 component spinor field. Substituting the
 partial wave decomposition into the action
 and integrating over the angular $\phi$, one can obtain
 \begin{eqnarray}
 S=\sum_m \int dt dr \frac{r}{2\pi} \sqrt{\frac{h}{f}}
 \psi_m^\dag(t,r)\gamma^0\left[\frac{\gamma^0}{\sqrt{h}}
 (\partial_t-imN^\phi)+
 \frac{\gamma^1}{\sqrt{f}}(\partial_{r*}-\frac{fh'}{4h})\right]\psi_m(t,r)\;.
 \end{eqnarray}
 Now our task is to interpret the reduced action (10)
 in terms of the two dimensional quantities.
 We consider the two dimensional metric
 \begin{eqnarray}
 ds^2=-h(r)dt^2+\frac{dr^2}{f(r)}\;.
 \end{eqnarray}
 In this background, we have calculated the
 spinor covariant derivative $D_\alpha^{(2)}$ of a two
 dimensional spinor field $\chi(t,r)$ contracted with
 two dimensional covariant gamma matrices $\sigma^i e_i^{(2)\alpha}$,
 which is explicitly given by
 \begin{eqnarray}
 \sigma^i e_i^{(2)\alpha} D_\alpha^{(2)}\chi=
 \left[\frac{\sigma^0}{\sqrt{h}}\partial_t
 +\frac{\sigma^1}{\sqrt{f}}\left(\partial_{r*}
 -\frac{fh'}{4h}\right)\right]\chi\;.
 \end{eqnarray}
 Now, we choose the following gamma matrices
 in three dimensional flat spacetime
 \begin{gather}
 \gamma^0=\begin{pmatrix}
 0 &1 \\
 -1 &0\end{pmatrix}\;,\;\;
 \gamma^1=\begin{pmatrix}
 0 &1 \\
 1 &0\end{pmatrix}\;,\;\;
 \gamma^2=\begin{pmatrix}
 1 &0 \\
 0 &-1\end{pmatrix}\;.
 \end{gather}
 While the gamma matrices in two dimensional flat spacetime
 are selected to be $\sigma^0=\gamma^0$ and $\sigma^1=\gamma^1$.
 We can rewrite the reduced action (10)
 in terms of the two dimensional quantities
 \begin{eqnarray}
 S=\sum_m\int dt dr \sqrt{-g^{(2)}} \Phi
 \bar{\psi}^{(2)}\sigma^i e_i^{(2)\alpha}
 \left(D_\alpha^{(2)}-imA_\alpha\right)\psi_m^{(2)}\;,
 \end{eqnarray}
 where the dilaton field $\Phi$ and gauge field $A_\alpha$
 associated with the charge $m$ are given by
 \begin{eqnarray}
 \Phi=\frac{r}{2\pi}\;,\;\;A_t=N^\phi\;,\;\;A_r=0\;.
 \end{eqnarray}

 It should be noted that the tetrad fields $e_a^\mu$
 we choose are convenient in the dimensional reduction
 process, but not the only one. One can get a new
 type of tetrad fields by performing a gauge
 rotation for the tetrad fields presented in Eq.(5),
 and use the new tetrad fields in calculations.
 If the corresponding two dimensional tetrad fields
 are properly selected, the form of the reduced two dimensional
 metric can be preserved.

 Up to now, using the dimensional reduction technique,
 we have found that the near-horizon physics for
 the fermionic field in the background
 of the general 2+1 dimensional stationary black hole
 can be approximated by an infinite collection of
 two component
 fermionic fields in 1+1 dimensional spacetime background
 coupled with the dilaton field and the $U(1)$
 gauge field.
 Therefore one can treat the original 2+1
 dimensional theory as a collection of
 $1+1$ dimensional quantum fields.

 Now the usual anomaly cancellation method
 can be applied to derive the Hawking radiation
 of spinor field.
 Since the event horizon is a null
 hypersurface, all ingoing modes
 at the horizon can not classically
 affect physics outside the horizon.
 Here, we focus on the effective theory outside
 the horizon and ignore the
 the classically irrelevant ingoing modes.
 Then, the theory becomes chiral and gauge or
 gravitational anomalies arise.
 If the symmetries of the theory are restored,
 these anomalies should be cancelled by the
 quantum effects of the classically irrelevant
 ingoing modes. It has been shown that the
 condition for anomaly cancellation
 can give rise to the
 Hawking flux of the charge, angular momentum
 and energy-momentum.

 In the present case,
 the original rotating symmetry in 2+1 dimensional
 spacetime is reduced to the gauge symmetry in two
 dimensional background.
 In the region near the horizon,
 the covariant current equations modified by Abelian anomaly
 is given by
 \begin{eqnarray}
 \nabla_\mu \widetilde{J}^\mu=-\frac{m^2}{4\pi\sqrt{-g^{(2)}}}
 \epsilon^{\mu\nu}F_{\mu\nu}
 \end{eqnarray}
 By solving this equation with appropriate boundary
 conditions that the covariant gauge currents vanish at the
 horizon, the flux of angular momentum from the horizon is given by
 \begin{eqnarray}
 F_a=-\frac{m^2}{2\pi}A_t(r_H)=\frac{m^2}{2\pi}\Omega_H
 \end{eqnarray}
 This is consistent with the flux derived
 from the Hawking distribution given by
 the Planck distribution with chemical
 potentials for angular momentums $m$
 of the fields radiated from the black hole,
 where the distribution for fermions
 is given by
 \begin{eqnarray}
 N_{m}=\frac{1}
 {e^{(\omega-m\Omega_H)/T_H}+1}\;.
 \end{eqnarray}

 Finally, we want to find the energy-momentum
 flux and the Hawking temperature of the black hole.
 The anomalous equation of the energy-momentum
 tensor in the region near the event horizon is given by
 \begin{eqnarray}
 \nabla_\mu T^\mu_{\;\;\nu}=F_{\mu\nu}
 \widetilde{J}^\mu+\frac{1}{\sqrt{-g^{(2)}}}
 \partial_\mu N^\mu_{\;\;\nu}\;,
 \end{eqnarray}
 where $N^\mu_{\;\;\nu}=\frac{1}{96\pi}
 \epsilon^{\beta\mu}\partial_\alpha\Gamma^{\alpha}_{\;\;\nu\beta}$.
 By applying the same process as in \cite{graanomaly1,graanomaly,investigations},
 the flux of the energy-momentum is determined as
 \begin{eqnarray}
 F_M&=&\frac{m^2}{4\pi}A_t^2(r_H)+N^r_t(r_H)\nonumber\\
 &=&\frac{m^2}{4\pi}\Omega_H^2
 +\frac{1}{192\pi}f'(r_H)h'(r_H)\;.
 \end{eqnarray}
 Comparing this result with
 the the energy-momentum flux derived from the
 Hawking distribution (18), which is given by
 \begin{eqnarray}
 F_M=\frac{m^2}{4\pi}\Omega_H^2
 +\frac{\pi}{12}T_H^2\;,
 \end{eqnarray}
 one can obtain the Hawking temperature as
 \begin{eqnarray}
 T_H=\frac{1}{4\pi}\sqrt{f'(r_H)h'(r_H)}\;,
 \end{eqnarray}
 which is consistent with the result calculated via the surface gravity.

 Up to now, we have succeeded in applying
 the quantum anomaly cancellation method
 to derive the Hawking radiation of fermionic field
 from the event horizon of the
 2+1 dimensional stationary black hole.
 This indicates that
 the method of anomaly cancellation to derive
 Hawking radiation is quite universal
 for different types of perturbative field.

 \section{Hawking radiation of fermionic
 field in BTZ black hole}

 In this section. we consider the BTZ black hole
 to give a demonstration of the
 general discussions in the last section.
 The BTZ black hole solution is an exact
 solution to Einstein field equation in a $2+1$ dimensional
 Einstein gravity with a negative cosmological
 constant $\Lambda=-1/l^2$ where the action is given by
 \begin{equation}
 I=\int dx^3 \sqrt{-g}(R+2\Lambda)\;.
 \end{equation}
 The BTZ black hole is described by the metric\cite{BTZ}
 \begin{equation}
 ds^2=-N^2
 dt^2+\frac{1}{N^2}dr^2+r^2(d\phi+N^{\phi}dt)^2\;,
 \end{equation}
 where
 \begin{equation}
 N^2=-M+\frac{r^2}{l^2}+\frac{J^2}{4r^2}\;,\;\;N^{\phi}=-\frac{J}{2r^2}\;,
 \end{equation}
 with $M$ and $J$ being the ADM mass and angular momentum
 of the BTZ black hole, respectively. The horizon
 is determined by the equation
 \begin{eqnarray}
 N^2=\frac{1}{l^2r^2}(r^2-r_+^2)(r^2-r_-^2)=0\;,
 \end{eqnarray}
 where $r_+$ and $r_-$ are locations
 of outer and inner horizons respectively.
 $r_+$ and $r_-$ are given by
 \begin{eqnarray}
 r_{\pm}^2=\frac{Ml^2}{2}\big[1\pm\sqrt{1-\frac{J^2}{M^2l^2}}\big]\;.
 \end{eqnarray}
 Comparing this metric with the general line element
 in the last section, one can easily find
 the near-horizon physics for
 the fermionic field
 in BTZ black hole
 can be approximated by an infinite collection of
 two component
 fermionic fields in 1+1 dimensional spacetime background
 coupled with the dilaton field and the $U(1)$
 gauge field, which are given by
 \begin{eqnarray}
 ds^2&=&-N^2dt^2+\frac{dr^2}{N^2}\;,\nonumber\\
 \Phi&=&\frac{r}{2\pi}\;,\;\;A_{t}=-\frac{J}{2r^2}\;,\;\;A_{r}=0\;.
 \end{eqnarray}
 Then, from eq.(17) and (20) the angular momentum flux and energy-momentum flux
 are respectively given by
 \begin{eqnarray}
 F_a&=&-\frac{m^2}{2\pi}A_t(r_H)=\frac{m^2}{4\pi}\frac{J}{r_+^2}\;,
 \nonumber\\
 F_M&=&\frac{m^2}{4\pi}A_t^2(r_H)
 +\frac{1}{192\pi}((N^2)')^2(r_H)
 \nonumber\\
 &=&\frac{m^2J^2}{16\pi r_+^4}+
 \frac{(r_+^2-r_-^2)^2}{48\pi l^4 r_+^2}\;,
 \end{eqnarray}
 from which the angular velocity and Hawking temperature
 can be read
 \begin{eqnarray}
 \Omega_H=\frac{J}{2r_+^2}\;,\;\;T_H=\frac{(r_+^2-r_-^2)}{2\pi l^2
 r_+}\;,
 \end{eqnarray}
 which is consistent with the
 results in the previous literatures\cite{li}.

 \section{Hawking radiation of fermionic
 field in the rotating black hole
 in BHT massive gravity}

 In this section, we will
 consider a rotating black hole
 solution in BHT massive gravity
 theory. BHT massive gravity
 theory is a new theory of
 massive gravity recently
 proposed by
 Bergshoeff, Hohm and Townsend\cite{BHT}.
 The theory is described by the action
 \begin{eqnarray}
 I_{BHT}=\frac{1}{16\pi G}\int
 d^3x \sqrt{-g}\left[R-2\lambda-
 \frac{1}{m^2}\left(R^{\mu\nu}R_{\mu\nu}
 -\frac{3}{8}R^2\right)\right]\;,
 \end{eqnarray}
 where $m$ is the mass parameter of the massive gravity
 and $\lambda$ is a constant which is different
 from the cosmological constant.
 The field equations are then of fourth order and given by
 \begin{eqnarray}
 G_{\mu\nu}+\lambda g_{\mu\nu}-\frac{1}{2m^2}K_{\mu\nu}=0\;,
 \end{eqnarray}
 where $G_{\mu\nu}=R_{\mu\nu}-\frac{1}{2}g_{\mu\nu}R$
 is the Einstein tensor and
 \begin{eqnarray}
 K_{\mu\nu}&=&2\nabla^2 R_{\mu\nu}-\frac{1}{2}\left(g_{\mu\nu}\nabla^2 R
 +\nabla_\mu\nabla_\nu R\right)
 -8R_{\mu\alpha}R_{\;\;\nu}^{\alpha}\nonumber\\
 &&+\frac{9}{2}RR_{\mu\nu}
 +g_{\mu\nu}\left(3R^{\alpha\beta}R_{\alpha\beta}
 -\frac{13}{8}R^2\right)\;.
 \end{eqnarray}

 In the special case, $m^2=\lambda=-\frac{1}{2l^2}$,
 the BHT massive gravity theory has the
 following rotating black hole solution
 \cite{BHTso}
 \begin{eqnarray}
 ds^2=-N(r)F(r)dt^2+\frac{dr^2}{F(r)}+r^2
 \left(d\phi+N^\phi(r) dt\right)^2\;,
 \end{eqnarray}
 where the metric functions are given by
 \begin{eqnarray}
 N&=&\left[1+\frac{bl^2}{4H}\left(1-\Xi^{1/2}\right)\right]^2\;,
 \nonumber\\
 N^\phi&=&-\frac{a}{2r^2}\left(4GM-bH\right)\;,
 \nonumber\\
 F&=&\frac{H^2}{r^2}\left[
 \frac{H^2}{l^2}+\frac{b}{2}\left(1+\Xi^{1/2}\right)H
 +\frac{b^2l^2}{16}\left(1-\Xi^{1/2}\right)^2
 -4GM\Xi^{1/2}
 \right]\;,
 \end{eqnarray}
 and
 \begin{eqnarray}
 H=\left[r^2-2GMl^2\left(1-\Xi^{1/2}\right)
 -\frac{b^2l^4}{16}\left(1-\Xi^{1/2}\right)^2
 \right]^{1/2}\;,
 \end{eqnarray}
 with
 \begin{eqnarray}
 \Xi=1-\frac{a^2}{l^2}\;.
 \end{eqnarray}

 The angular momentum is given by $J = Ma$, where $M$ is the
 mass measured with respect to the zero mass black hole
 and $l<a<l$ is the rotation parameter.
 Except for the mass $M$ and the rotation parameter $a$,
 the solution is also described by
 an additional1 gravitational hair parameter $b$.
 The horizon is determined by the equation $F=0$,
 which can be solved as
 \begin{eqnarray}
 r_H=\frac{l^2}{2}\sqrt{\frac{1+\Xi^{1/2}}{2}}
 \left(\sqrt{b^2+\frac{16GM}{l^2}}-b\Xi^{1/4}
 \right)\;.
 \end{eqnarray}

 Comparing the metric with the general line element
 in the section II, one can find
 the near-horizon physics for
 fermionic field in this black hole solution
 is approximated by an infinite collection of
 two component fermionic fields in 1+1
 dimensional spacetime background
 coupled with the dilaton field and the $U(1)$
 gauge field, which are given by
 \begin{eqnarray}
 ds^2&=&-NFdt^2+\frac{dr^2}{F}\;,\nonumber\\
 \Phi&=&\frac{r}{2\pi}\;,\;\;
 A_{t}=-\frac{a}{2r^2}\left(4GM-bH\right)\;,\;\;A_{r}=0\;.
 \end{eqnarray}
 Then, from eq.(17) and (20) one can find
 the angular momentum flux and energy-momentum flux
 which are respectively given by
 \begin{eqnarray}
 F_a&=&-\frac{m^2}{2\pi}A_t(r_H)
 =\frac{m^2}{2\pi}\frac{\left(1-\Xi^{1/2}\right)}{a}\;,
 \nonumber\\
 F_M&=&\frac{m^2}{4\pi}A_t^2(r_H)
 +\frac{1}{192\pi}(NF)'F'\big|_{r_H}
 \nonumber\\
 &=&\frac{m^2}{4\pi}\frac{\left(1-\Xi^{1/2}\right)^2}{a^2}+
 \frac{1}{96\pi}\Xi
 \left(b^2+\frac{16GM}{l^2}\right)
 \left(1+\Xi^{1/2}\right)^{-1}\;,
 \end{eqnarray}
 from which the angular velocity and Hawking temperature
 can be read
 \begin{eqnarray}
 \Omega_H&=&-A_t(r_H)=\frac{\left(1-\Xi^{1/2}\right)}{a}\;,
 \nonumber\\
 T_H&=&\frac{1}{4\pi}(NF)'F'\big|_{r_H}=\frac{1}{2\pi}\sqrt{\frac{\Xi}{2}
 \left(b^2+\frac{16GM}{l^2}\right)
 \left(1+\Xi^{1/2}\right)^{-1}}\;.
 \end{eqnarray}
 This is consistent with the results
 given in \cite{BHTso}.

 \section{conclusion}

 In this paper, the method of anomaly cancellation
 to derive Hawking radiation
 initiated by Robinson and Wilczek is applied to
 2+1 dimensional stationary black holes.
 In particular, we have found
 the anomaly method can be used to
 investigate the Hawking radiation
 of spin $\frac{1}{2}$ field,
 which indicates the anomaly method
 is quite universal.

 The most essential observation
 in this paper is that
 the near-horizon physics for
 the spin $\frac{1}{2}$ field in the background
 of the general 2+1 dimensional stationary black hole
 can be approximated by an infinite collection of
 two component
 fermionic fields in 1+1 dimensional spacetime background
 coupled with the dilaton field and the $U(1)$
 gauge field.
 This permits us to derive Hawking radiation
 of fermionic field
 by solving the anomaly current equation for gauge field and
 energy-momentum tensor in the reduced
 two dimensional background.
 As an example, we also apply this method
 to two types of black holes in three dimensional
 spacetime, which are BTZ black hole in Einstein gravity
 and a rotating black hole
 in Bergshoeff-Hohm-Townsend (BHT) massive gravity.
 It is shown that the Hawking temperatures are recovered and they are
 consistent with the results obtained previously
 in literatures.

 \section*{ACKNOWLEDGEMENT}

 This work wa
s supported by the National Natural Science Foundation
 of China and Cuiying Project of Lanzhou University.

 \end{document}